\documentstyle[prl,aps,floats,epsf,color]{revtex}
\begin{document}
\baselineskip=12pt
\def\be{\begin{equation}}
\def\ee{\end{equation}}
\def\bea{\begin{eqnarray}}
\def\eea{\end{eqnarray}}
\def\E{{\rm e}}
\def\bearst{\begin{eqnarray*}}
\def\eearst{\end{eqnarray*}}
\def\peleven{\parbox{11cm}}
\def\peffec{\peight{\bearst\eearst}\hfill\peleven}
\def\pspace{\peight{\bearst\eearst}\hfill}
\def\ptwelve{\parbox{12cm}}
\def\peight{\parbox{8mm}}
\twocolumn[\hsize\textwidth\columnwidth\hsize\csname@twocolumnfalse\endcsname

\title
{Fractal Analysis of Discharge Current Fluctuations }

\author{S. Kimiagar$^1$, M. Sadegh Movahed$^{2}$, S. Khorram$^{3}$, S. Sobhanian$^{4,5}$, and M. Reza Rahimi Tabar$^{6,7,8}$ }
\address{$^{1}$Department of Physics,science faculty, Tehran central branch, Islamic Azad University, Tehran,
Iran}
\address{$^{2}$Department of Physics, Shahid Beheshti university, G.C., Evin, Tehran 19839, Iran}
%\address{$^{3}$ School of Astronomy, Institute for Studies in theoretical Physics and Mathematics, P.O.Box 19395-5531,Tehran,
%Iran}
\address{$^{3}$ Research Institute for Applied Physics and Astronomi, University of Tabriz, Tabriz 51664, Iran}
\address{$^{4}$ Department of Theoretical Physics and Astrophysics, Tabriz
University, Tabriz 51664, Iran}
\address{$^{5}$ Research Institute for Fundamental Sciences, Tabriz
51664, Iran}
\address{$^{6}$Department of Physics, Sharif University of
Technology, P.O.Box 11365--9161, Tehran, Iran}
\address{$^{7}$CNRS
UMR 6529, Observatoire de la C$\hat o$te d'Azur, BP 4229, 06304 Nice
Cedex 4, France \\ $^8$Carl von Ossietzky University, Institute of
Physics, D-26111 Oldenburg, Germany }

\vskip 1cm

 \maketitle
%\date{00/07/2000}
%\maketitle

%%%%%%%%%%%%%%%%%%%%%%%%%%%%%%%%%%%%%%%%%%%%%%%%%%%%%%
%ABSTRACT
%%%%%%%%%%%%%%%%%%%%%%%%%%%%%%%%%%%%%%%%%%%%%%%%%%%%%%

\begin{abstract}

We use the multifractal detrended fluctuation analysis (MF-DFA) to
study the electrical discharge current fluctuations in plasma and
show that it has multifractal properties and behaves as a weak
anti-correlated process. Comparison of the MF-DFA results for the
original series with those for the shuffled and surrogate series
shows that correlation of the fluctuations is responsible for
multifractal nature of the electrical discharge current.\\

%PACS numbers: 05.45.-a, 05.45.Df, 47.53.+n, 52.35.Ra
 \end{abstract}
\hspace{.3in}
\newpage
 ]
\section{Introduction}

Plasma physics is concerned with the complex interaction of many
charged particles with external or self-generated electromagnetic
fields. It plays an essential role in many applications, ranging
from advanced lighting devices to surface treatments for
semiconductor applications or surface layers. At the same time, the
interpretation and estimation of physical and chemical properties of
a plasma fluid have been one of the main research areas in the
science of magnetohydrodynamics and transport theory
\cite{1,2,3,4,5,6,7,bud,zaj,van,bud1,bud2,wei,car1,dink05}. As in
many other fields in physics, the complex physics requires advanced
numerical tools to be developed and used.

It is well-known that the discharge current fluctuations in the
plasma often exhibit irregular and complex behavior. Indeed, the
current fluctuations represent a dynamical system influenced by many
factors, such as the pressure, the electrical potential between
cathode and anode, the electrical properties of the gas, noises, and
trends, due to the experimental setup, etc. Factors that influence
the trajectory of discharge current fluctuations have enormously
large phase space. Thus, the use of stochastic tools for
investigating their statistical properties is natural. Because of
the complexity and stochasticity of the discharge fluctuations in
the plasma fluid, it is generally difficult to have access to
detailed dynamics of the plasma ions, without paying attention to
the statistical aspect of plasma. Therefore, there may be no remedy,
except using stochastic analysis to investigate the evolution and
physical properties of the discharge current produced by such
fluctuations. Also due to the limitations in the experimental setup
for measuring the fluctuations, as well as the finiteness of the
available data in some cases, the original fluctuations may be
affected by some trends and non-stationarities. Therefore, in order
to infer valuable statistical properties of the original
fluctuations and avoid spurious detection of correlations, one must
use a robust method which should be insensitive to any trends.

Fluctuation of the electric and magnetic fields of plasma, spectral
density, logistic mapping and nonlinearity of ionization wave have
been investigated in
\cite{1,2,3,4,5,6,7,bud,zaj,van,bud1,bud2,wei,car1}. Recently,
Carreras {\it et. al.}, have shown that the plasma has a
multifractal nature with intermittency levels comparable to the
levels measured in neutral fluid turbulence\cite{car1}.
%have investigated plasma fluctuations by
%notice to degree of intermittency and have shown that plasma
%fluctuations have a multifractal nature with intermittency levels
%comparable to the levels measured in neutral fluid
%turbulence\cite{car1}.
Also, Budaev {\it et. al.}, used the scaling behavior of structure
functions and wavelet transform modulus maxima (WTMM). They showed
the anomalous transport of particles in the plasma phase attached to
the turbulent property has multifractal nature
\cite{bud,bud1,bud2,wei}.
%by using the scaling behavior of
%structure function and wavelet transform modulus maxima (WTMM)
%showed that the anomalous transport of particles in the plasma phase
%attached to the turbulent property has multifractal nature
%\cite{bud,bud1,bud2,wei}.

 Although the analysis of discharge
current in the plasma has a long history \cite{sig02,isu05,Den06},
nevertheless, some important issues, such as ions and electrons
acceleration mechanism, interaction between laser and plasma,
especially from the statistical properties point of view, fractal
features, effects of trends in small and large scales and the kind
of correlations have remained unexplained
\cite{1,2,3,4,5,6,7,bud,zaj,van,bud1,bud2,wei,car1,dink05,sig02,isu05}.

Generally, correlated and uncorrelated time series could have same
probability distribution function. Also they may have mono-fractal
or multi-fractal nature. The mono-fractal signals can be describe by
one scaling exponent.
% and  the exponent of second moment will give
%all of the high moment exponents by simple linear relation.
However many time series do not exhibit a simple monofractal scaling
behavior. In some cases, there exist crossover (time-) scales
separating regimes with different scaling exponents. In other cases,
the scaling behavior is more complicated, and different scaling
exponents are required for different parts of the series. This
occurs, e. g., when the scaling behavior in the first half of the
series differs from the scaling behavior in the second half. In even
more complicated cases, such different scaling behavior can be
observed for many interwoven fractal subsets of the time series. In
this case a multitude of scaling exponents is required for a full
description of the scaling behavior, and a multifractal analysis
must be applied. In nature, two different types of multifractality
in time series can be distinguished: (i) Multifractality due to a
broad probability density function for fluctuations. In this case
the multifractality cannot be removed by shuffling the series. (ii)
Multifractality due to different (time-) correlations for small and
large fluctuations. %In this case the surrogate of time series ( to
%make it Guassian) can not affect the multifractality.
If both kinds of multifractality are present, the shuffled series
will show weaker multifractality than the original series.

\begin{figure}[t]
\epsfxsize=8truecm\epsfbox{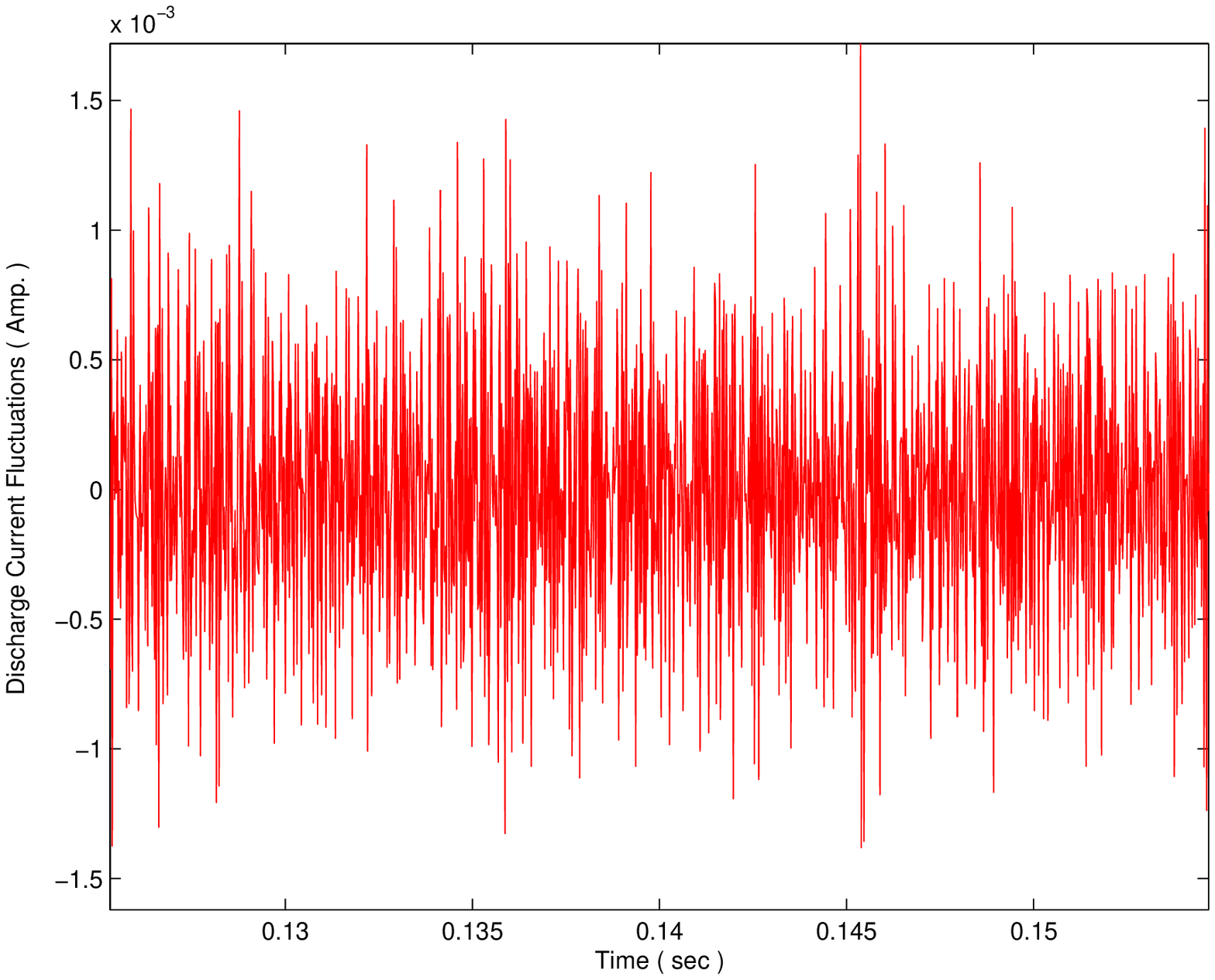} \narrowtext
\epsfxsize=8truecm\epsfbox{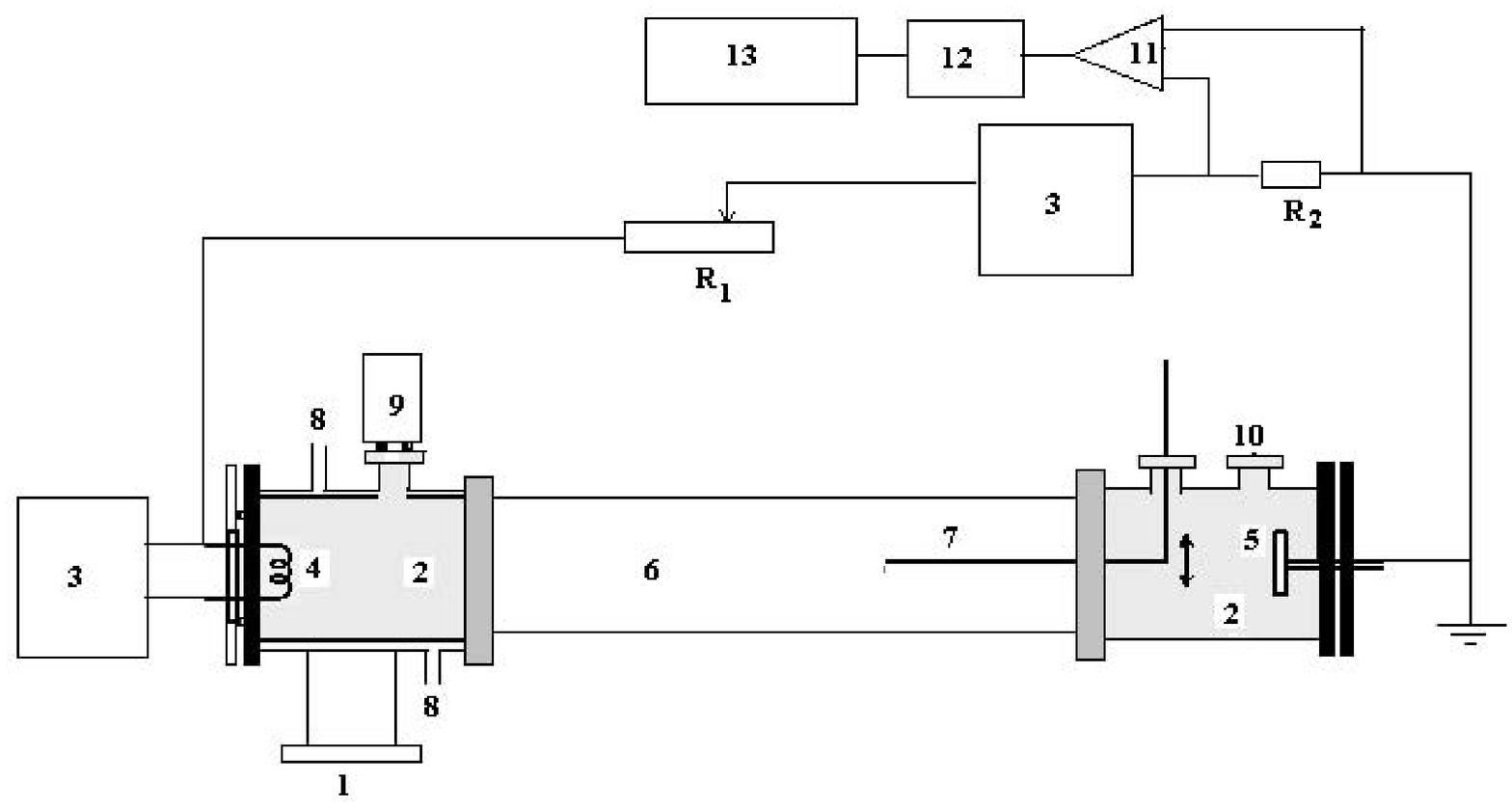} \narrowtext \caption{Upper
panel: typical discharge current fluctuations as a function of time
in our experimental setup. Lower panel: the sketch of the
experimental setup to record the discharge current fluctuations in
the tube, filled by helium, with (1) vacuum pumps; (2) copper
cylinder; (3) power supply; (4) hot cathode; (5) anode plate; (6)
glass tube; (7) single Langmuir probe; (8) water cooling; (9) Pirani
pressure gage; (10) gas inlet; (11) OP-Amp; (12) low pass filter,
and (13) A/D card and PC.}\label{fig1}
\end{figure}

Here we rely on the state-of-the-art of computational methods in
statistical physics to characterize the complex behavior of
electrical discharge time series. We study the discharge current
fluctuations (see the upper panel of Figure \ref{fig1}) by the
multifractal detrended fluctuation analysis (MF-DFA) and
Fourier-detrended fluctuation analysis (F-DFA) methods. Using the
method proposed in \cite{kunhu,trend2}, we investigate the relation
between the amplitude and the period of the trend and crossover in
discharge current fluctuations in the plasma.

The rest of this paper is organized as follows: In Section II we
describe some important steps of fractal analysis to explore the
stochastic time series in the presence of sinusoidal trends. The
Hurst exponent and its relation to the classical multifractal, the
generalized multifractal dimension and the H\"older exponents are
described in section II. Data preparation, experimental setup,
surrogate and shuffled time series are also given in section II. We
then eliminate the sinusoidal trends via the F-DFA technique in
section III, and investigate the multifractal properties of the
remaining fluctuations. In section IV we deal with the source of
multifractality in data. Section V is devoted to a summary of the
results.

%The description of the data and the experimental setup to measure
%the discharge fluctuations in the plasma of helium companying the
%method for making the surrogate and shuffled data are also described
%in section II. We then eliminate the sinusoidal trends via the F-DFA
%technique in section III, and investigate the multifractal
%properties of the remaining fluctuations. In section IV we deal with
%the source of multifractality in data. For this purpose we compare
%the MF-DFA results for the remaining data set with the results
%obtained via the MF-DFA for the shuffled and surrogate series.
%Section V is devoted to a summary of the results.

%%%%%%%%%%%%%%%%%%%%%%%%%%%%%%%%%%%%%%%%%%%%%%%%%%%%%%%%%%%%%%%%%%%%%%%%%
%%%%%%%%%%%%%%%%%%%%%%%%%%%%%%%%%%%%%%%%%%%%%%%%%%%%%%%%%%%%%%%%%
%%%

\section{Analysis techniques and Experimental setup }

The simplest type of multifractal analysis is based upon the
standard partition function, which has been developed for
multifractal characterization of normalized, stationary measurements
\cite{feder88,barabasi,peitgen,bacry01}. The standard formalism does
not yield the correct results for nonstationary time series that are
affected by trends. The improved multifractal formalism  has been
developed by Muzy {\it et al.} \cite{muzy}, which is known as the
wavelet-transform modulus maxima (WTMM) method \cite{muzy,wtmm}.
%Thus,
%systematically improved multifractal formalism was first has been
%developed by Muzy {\it et al.} \cite{muzy}, which is known as the
%wavelet-transform modulus maxima (WTMM) method \cite{muzy,wtmm}.
 It is based on the wavelet analysis and involves tracing the maxima
lines in the continuous wavelet transform over all the scales. The
second method which is known as MF-DFA, is based on the
identification of the scaling behavior of the $q$th moments and is
the generalization of the standard DFA which uses only the second
moment, $q=2$
\cite{kunhu,trend2,Peng94,Ossa94,murad,physa,trend3,trend4}.

%
%The second method, the MF-DFA, is based on the identification of the
%scaling of the $q$th moments, depending on the series length, and is
%the generalization of the standard DFA which uses only the second
%moment, $q=2$. The MF-DFA methods are the modified version of the
%DFA for detecting multifractal properties of time series
%\cite{kunhu,trend2,Peng94,Ossa94,murad,physa,trend3,trend4}.

It has successfully been applied to diverse problems, such as heart
rate dynamics \cite{Peng95,herz,PRL00}, economical time series
\cite{economics,economics1,economics2,sadeghmarket,Ivanov04,sadegheco},
river flow \cite{sadeghriver} and sunspot fluctuations
\cite{sadeghsun}, cosmic microwave background radiations
\cite{sadeghcmb}, and music \cite{jafarimusic1,jafarimusic2,music3}.

In general, experimental data are often affected by
non-stationarities, such as trends which must be well-distinguished
from the intrinsic fluctuations of the series, in order to determine
their correct scaling behavior. In addition, very often we do not
know the reasons for the underlying trends in the collected data
and, even worse, we do not know the scales of the underlying trends.
For reliable detection of the correlations, it is essential to
distinguish trends from the intrinsic fluctuations in data. Hurst
rescaled-range analysis \cite{hurst65} and other non-detrending
methods work well if the records are long and do not involve trends.
But, if trends are present in the data, they might yield inaccurate
results. In general, the MF-DFA is a well-established method for
determining the scaling behavior of noisy data in the presence of
trends without knowing their origin and shape
\cite{Peng94,Ossa94,Peng95,allan,buldy95}. Here in order to
eliminate the effect of sinusoidal trend, we apply the Fourier DFA
(F-DFA) \cite{na04,chi05}. After elimination of the trend we use the
MF-DFA to analyze the data set.

\subsection{MF-DFA Method}

The MF-DFA consists of the following four steps (see
\cite{kunhu,trend2,Peng94,murad,physa,trend3,trend4,bun02} for more
details):

(i): Computing the profile of underlying data series, $x_k$, as
\begin{equation}
Y(i) \equiv \sum_{k=1}^i \left[ x_k - \langle x \rangle \right]
\qquad i=1,\ldots,N \label{profile}
\end{equation}
(ii): Dividing the profile into $N_s \equiv {\rm int}(N/s)$
non-overlapping segments of equal lengths $s$, and then computing
the fluctuation function for each segment
\begin{equation}
F^2(s,m) \equiv {1 \over s} \sum_{i=1}^{s} \left\{ Y[(m-1) s + i] -
y_{m}(i) \right\}^2 \label{fsdef}
\end{equation}
where $y_{m}(i)$ is a fitting polynomial in segment $m$th. Usually,
a linear function is selected for fitting the function. If there do
not exist any trends in the data, a zeroth-order fitting function
might be enough \cite{Peng94,Ossa94,PRL00}.\\
 (iii): Averaging the local fluctuation function over all the part, given
by
\begin{equation}
F_q(s) \equiv \left\{ {1 \over  N_s} \sum_{m=1}^{ N_s} \left[
F^2(s,m) \right]^{q/2} \right\}^{1/q} \label{fdef}
\end{equation}

Generally, $q$ can take any real value, except zero. For $q=2$, the
standard DFA procedure is retrieved.\\
(iv): The final step is determining the slope of the log-log plot of
$F_q(s)$ versus $s$ directly determines the so-called generalized
Hurst exponent $h(q)$, as
\begin{equation}
F_q(s) \sim s^{h(q)} \label{Hq}
\end{equation}

For stationary time series, such as the fractional Gaussian noise
(fGn), $Y(i)$ in Eq. (\ref{profile}) will be a fractional Brownian
motion (fBm), and so, $0<h(q=2)<1.0$. The exponent $h(2)$ is
identical with the well-known Hurst exponent $H$
\cite{feder88,Peng94,Ossa94,murad}. Moreover, for a nonstationary
series, such as the fBm, $Y(i)$ in Eq. (\ref{profile}) will be a sum
of the fBm series and, thus, the corresponding scaling exponent of
$F_q(s)$ is identified by $h(q=2)>1.0$ \cite{Peng94,Ossa94,eke02}
(see the appendix of \cite{sadeghriver,sadeghsun} for more details).
In this case, the relation between the exponent $h(2)$ and $H$ is
$H=h(q=2)-1$. The auto-correlation function is characterized by a
power law, $C(s)\equiv\langle n_kn_{k+s} \rangle \sim s^{-\gamma}$,
with $\gamma=2-2H$. Its power spectrum is given by,
$S(\nu)\sim\nu^{-\beta}$, with frequency $\nu$ and $\beta=2H-1$. In
the nonstationary case, the correlation function is
\begin{equation}
C(i,j)=\langle x_ix_{j}\rangle\sim i^{2H}+j^{2H}-|i-j|^{2H}
\end{equation}
where $i,j\geq 1$ and the power-spectrum scaling exponent is,
$\beta=2H+1$ \cite{Peng94,Ossa94,sadeghriver,sadeghsun,eke02}.

 For
monofractal time series, $h(q)$ is independent of $q$, since the
scaling behavior of the variances $F^2(s,m)$ is identical for all
the segments $m$, and the averaging procedure in Eq. (\ref{fdef})
will just yield identical scaling behavior for all values of $q$. If
we consider positive values of $q$, the segments $m$ with large
variance $F^2(s,m)$ (i.e., large deviations from the corresponding
fit) will dominate the average $F_q(s)$.  Thus, for positive values
of $q$, $h(q)$ describes the scaling behavior of the segments with
large fluctuations. For negative values of $q$, on the other hand,
the segments $m$ with small variance $F^2(s,m)$ will dominate in the
average $F_q(s)$, and $h(q)$ describes the scaling behavior of the
segments with small fluctuations.

The classical multifractal scaling exponents $\tau(q)$, defined by
the standard partition function-based formalism, discussed in
literature
\cite{feder88,barabasi,peitgen,bacry01,physa,trend3,trend4,sadeghsun},
is related to the generalized hurst exponent via the MF-DFA as
\begin{equation}
\tau(q)=q h(q) - 1 \label{tauH}
\end{equation}
Moreover, the generalized multifractal dimensions $D(q)$ read as
\begin{equation}
D(q) \equiv {\tau(q) \over q-1}={qh(q)-1\over q-1}\label{Dq}
\end{equation}

Another way of characterizing a multifractal series is through its
singularity spectrum $f(\alpha)$, which is related to $\tau(q)$ via
a Legendre transform \cite{feder88,peitgen}. Here, $\alpha$ is the
singularity strength or the H\"older exponent. Using Eq.
(\ref{tauH}), we can directly relate $\alpha$ and $f(\alpha)$ to
$h(q)$
\begin{equation}
\alpha =h(q) + q h'(q) \quad {\rm and} \quad
f(\alpha)=q[\alpha-h(q)]+1 \label{Legendre2}
\end{equation}

A single H\"older exponent denotes monofractality, while in the
multifractal case, different parts of the structure are
characterized by different values of $\alpha$, leading to the
existence of the spectrum $f(\alpha)$.

In some cases, there exist one or more crossover (time) scales,
$s_\times$, segregating regimes with different scaling exponents,
e.g., the correlation exponent for $s\ll s_{\times}$ and another
type of the correlation or uncorrelated behavior for $s\gg
s_{\times}$ \cite{kunhu,trend2,physa,trend3,trend4}. In the presence
of different behavior of the various moments in the MF-DFA method,
distinct scaling exponents are required for different parts of the
series \cite{kunhu,trend2}. Therefore, one needs a multitude of
scaling exponents (multifractality) for a full description of the
scaling behavior. A crossover usually can arise from a change in the
correlation properties of the signal at different time or space
scales, or can often arise from trends in the data
\cite{kunhu,trend2}. However it is well-known WTMM method can remove
this crossover but, in many case, the presence of crossover as well
as their values have physical importance. %, so that we have to use
%those methods which are sensitive to the kind of trend in one hand
%and on the other hands one should manipulate the effect of each
%sinusoidal trends embedded in data set on the crossover scales.

Let us mention two advantages of the MF-DFA method in order to
compare with the applicability of WTMM method. The first advantage
corresponds to the effort of programming and second one is related
to the performance and reliability of given results. The MF-DFA does
not require the modulus maxima procedure while WTMM need to do this
task. The wavelet coefficients can become arbitrary small in WTMM
method. Subsequently MF-DFA does not involve more effort in
programming as well as more time consuming in contrast to WTMM,
specially for long length time series such as our underlying data
set \cite{bun02}. The second advantage is related to the fact that
the MF-DFA gives more reliable results than WTMM specially, for
negative moments. It has been reported that WTMM gives an
overestimated multifractal exponent and in some cases WTMM can also
give different results if one applies different wavelets
\cite{paw05}. However, the most disadvantage and limitation of
MF-DFA method will appear when it is applied to investigate data set
with small size. In this case MF-DFA gives rich singularity spectrum
corresponds to more multifractality property than what should be
existed. Fortunately this circumstance does not occur in our
situation, because the typical size of our plasma data is about
$10^6$.

In order to remove the trends correspond to the low frequency
periodic behavior, we transform the recorded data to the Fourier
space using the method proposed in \cite{cooly65} (see also
\cite{kunhu,trend2,physa,trend3,trend4}). Using this method we can
track the influence of sinusoidal trends on the results and
determine the value of so-called crossover in the fluctuation
function, in terms of the scale in DFA method. We determine over
which scale noises or trends have dominant contribution
\cite{na04,chi05,koscielny98,koscielny98b}. After removing the
dominant periodic functions, such as sinusoidal trends, we obtain
the fluctuation exponent by direct application of the MF-DFA. If
truncation of the number of the modes be sufficient, the crossover
due to a sinusoidal trend in the log-log plot of $F_q(s)$ versus $s$
disappear.

\subsection{Data description and experimental setup}

To investigate the stochastic nature of the discharge current
fluctuations in a typical plasma, we constructed an experimental
setup shown in the lower panel of Figure \ref{fig1}. The discharge
glass tube has two copper ends, $80$ mm in diameter and 110 cm in
length. One end is the anode electrode (a flat copper plate as a
positive pole), while the other end represents the cathode (tungsten
filament as a negative pole and electron propagator). The discharge
tube is evacuated to a base pressure of $0.1$ up to $0.8$ torr under
a voltage of $400 - 900$V and filled with Helium as the working gas.
The pressure, voltage and current should be optimal for ensuring the
stability of the plasma. The discharge current fluctuations were
monitored using a resistor which was connected to an operational
amplifier impedance converter. We fixed the pressure and
investigated how the statistical properties of plasma changing under
variation of the current. The fluctuations of the discharge current
were digitized and cleaned with a filter that omitted direct
current. Thereafter, the fluctuation of the discharge were recorded
at a rate of $44100$ $sample/sec$, with a resolution of 12 bits,
using a analog-digital card for several values of the electrical
discharge current intensity, namely, $50, 60, 100, 120, 140, 180$,
and $210$ mA. The typical size of the recorded data  for every
current intensity is about $10^6$.

\subsection{Surrogate and shuffled data }

The phase-randomized surrogate consists of three steps
\cite{dean,panter,sur1,sur2}:

(i) Computing the discrete fourier transform (DFT) coefficients of
the series
\begin{equation}
{\mathcal{F}}^2\{x(t)\}\equiv\left|X(\nu)\right|^2=\left|X(k)\right|^2=\left|\frac{1}{\sqrt{N}}\sum_{n=0}^{N-1}x(t_n)e^{i2\pi
nk/N}\right|^2
\end{equation}
where $\nu=k/N\Delta t$ and $\Delta t$ is the step of digitalization
in the experimental setup.

(ii) Multiplying the DFT coefficients of the series by a set of
pseudo-independent, uniformly distributed $\phi(\nu)$ quantities in
the range $[0,2\pi)$
\begin{equation}
\tilde{X}(\nu)=X(\nu)e^{i\phi(\nu)}
\end{equation}

(iii) The surrogate data set is given by the inverse DFT as

\begin{equation}
{\mathcal{F}}^{-1}\{{\tilde{X}}(\nu)\}\equiv\tilde{x}(t_n)=\frac{1}{\sqrt{N}}\sum_{k=0}^{N-1}\left|X_k\right|e^{i\phi(k)}e^{-i2\pi
nk/N}
\end{equation}

The power spectrum of the surrogate data set is the same as one for
the original data. According to the Wiener-Khintchine theorem,
surrogate data has the same autocorrelation as the initial series
\cite{dean,panter}. The amplitude of the surrogate data will be
preserved as in the original data. However the probability density
function of the data will change to the Gaussian distribution. We
note that applying the DFT needs the data to be periodic
\cite{galka}. In addition, this procedure eliminates nonlinearities,
preserving only the linear properties of the underlying original
data set \cite{panter}.

To produce a shuffled data set, one should clean the imposed memory
in series. To this end, we should randomize the order of data in
underlying series while their values remain unchanged.

\section{Fractal analysis of electrical discharge fluctuations time series} \label{detcross}

As mentioned in section II, spurious correlations may be detected if
the time series be nonstationary, or is affected by trends. In such
cases, direct calculation of the correlation exponent, the spectral
density, the fractal dimensions, etc., do not yield reliable
results. Our data sets are affected by some trends, such as the
alternative current oscillation, the noise due to the electronic
instruments, and the fluctuations of striation areas near the anode
and cathode plates. Therefore, we must use detrended methods to
distinguish the intrinsic fluctuations from the nonstationarity and
trends.

\begin{figure}[t]
\epsfxsize=8.2truecm\epsfbox{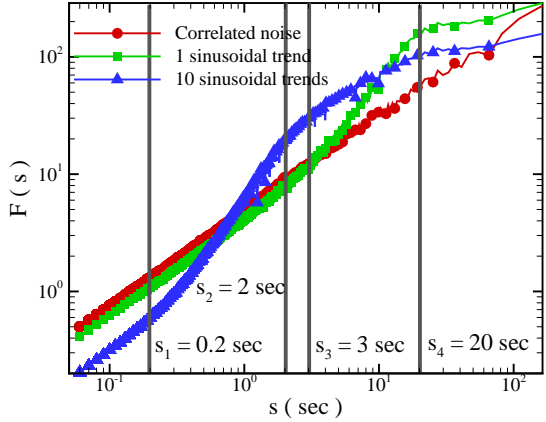}\narrowtext
\epsfxsize=8.2truecm\epsfbox{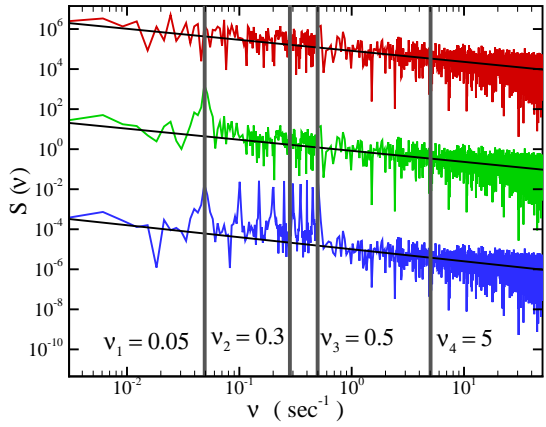} \caption{Upper panel shows
the fluctuations function for the correlated noise and its
superposition with one and ten sinusoidal trends. Lower panel
corresponds to their power spectrum. The curves, from top to bottom,
correspond to the correlated noise, one and ten embedded sinusoidal
trends in the noise, respectively. Oblique solid lines correspond to
the scaling behavior of power spectrum of the clean correlated
noise, namely, $\beta=0.6$ in $S(\nu)\sim \nu^{-\beta}$.}
\label{sin}
\end{figure}

Let us determine whether the data set has a sinusoidal trend or not.
According to the MF-DFA method, the generalized Hurst exponents
$h(q)$ in Eq. (\ref{Hq}) are determined by analyzing the log-log
plots of $F_q(s)$ versus $s$ for each $q$. Using the rate of
digitization in the experimental setup, $44100$ $sample/sec$, one
can simply change the unit of $s$ to seconds. It must be pointed out
that, to infer the desired exponents and to avoid errors arising for
small values of $s$ \cite{kunhu,trend2,physa,trend3,trend4}, we use
the interval $s\geq 0.005$ sec in our analysis. The resolution of
the recorded data in our setup is $1/44100\sim 0.00023$ sec. We use
this interval throughout the paper, unless specified otherwise. Our
investigation indicates that there is at least one crossover time
scale in the log-log plots of $F_q(s)$ versus $s$ for every $q$. To
determine its value, we use the following two criteria and combine
their results:

(i) Based on the recent results by \cite{kunhu,trend2} and
\cite{sadeghriver,sadeghsun}, every sinusoidal trend in the data
causes some crossovers in the scaling function, $F_q(s)$, derived by
the MF-DFA. The number of such crossovers depends on the size of
data and the wavelength of the sinusoidal trends
\cite{kunhu,trend2}. It is well known that the crossovers divide the
fluctuations function into some regions that correspond to various
scaling behaviors of $F_q(s)$ versus $s$, that are related to the
competition between noise and trends \cite{kunhu,trend2}. To prove
this statement and show how one can determine the value of
crossovers, we generated numerically a time series which is a
superposition of a correlated noise, namely, with the Hurst exponent
$H=0.8$, and a sinusoidal trend with its period equal to $T=20$ sec.
As shown in Figure \ref{sin}, one sinusoidal trend is embedded in
the data, and there are two crossovers at $s_3 \sim 3$ sec and $s_4
\sim 20$ sec in the fluctuation function. The larger crossover is
directly associated with the period of the sinusoidal trend
\cite{kunhu,trend2}.

The crossovers are also confirmed by resorting to the power spectrum
of the data shown in the middle graph of the lower panel of Figure
\ref{sin}. One can observe that for $\nu>\nu_2=1/s_3\sim 0.3$
sec$^{-1}$, the scaling behavior of the power spectrum is the same
as those for the correlated noise, indicating that in these scales
the noise effect is dominant. Consequently, the scaling behavior of
$F_q(s)$ for $s<s_3$ is very close to those of the correlated noise
alone (see the upper panel of Figure \ref{sin}).

\begin{figure}[t]
\epsfxsize=8.2truecm\epsfbox{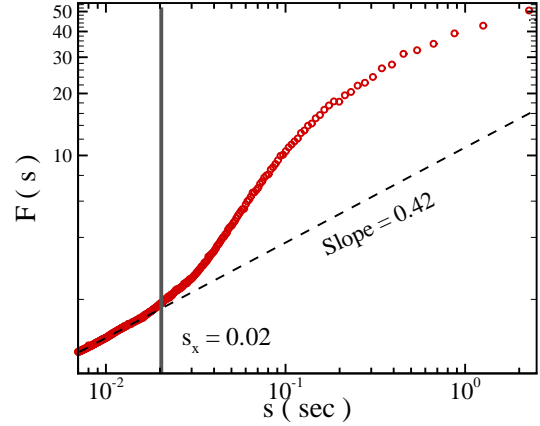}\narrowtext
\epsfxsize=7.2truecm\epsfbox{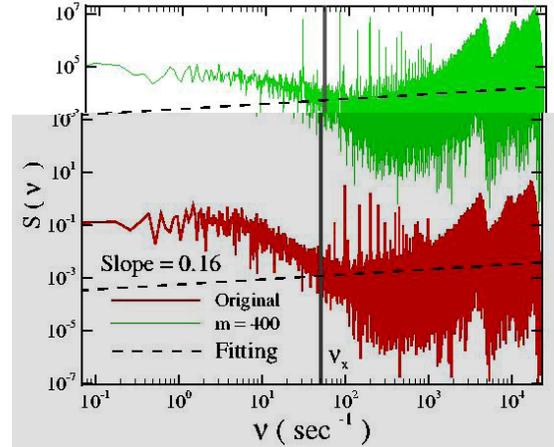} \caption{Upper panel shows
the crossover behavior of the log-log plot of $F(s)$ versus $s$ for
the original time series for $q=2.0$. Lower panel corresponds to the
power spectrum of the original and cleaned data set. Oblique solid
lines correspond to the scaling behavior of the power spectrum of
the cleaned series, namely, $\beta=-0.16$ in $S(\nu)\sim
\nu^{-\beta}$.} \label{crossover}
\end{figure}

Thereafter, we embedded ten sinusoidal trends with various
frequencies in the original signal, and performed the same
computations as shown in Figure \ref{sin}. We found that, when one
increases the number of sinusoidal trends in the original noise,
then  expects that the value of the crossover at large scales,
namely, $s_4(\nu_1)$ ( which is related to the dominant embedded
sinusoidal trends), will extend, as shown in Figure \ref{sin}. In
other words, we have an interval, i.e., $s\in [s_2,s_4]$, or in
frequency space, $\nu\in[\nu_1,\nu_3]$, within which the scaling
behavior of the fluctuation function changes smoothly. Therefore, we
cannot determine an exact value for the crossover at large scales.
Moreover, as shown in the upper panel of Figure \ref{sin}, at time
scales smaller than $s_2$ or for frequencies larger than $\nu_3$,
the fluctuation function retrieves its noisy behavior. This
observation depends on the amplitudes and frequencies of the
embedded periodic trends. Therefore, the exact value of the
crossover at small scale is not obvious in the spectral density and,
hence, we use the following criteria.

ii) As discussed above the existence of many sinusoidal trends in
our data set, we expect that a plot of $F_q(s)$ versus $s$ possess
at least one crossover. This crossover divides $F_q(s)$ into two
regions, as shown in the upper panel of Figure \ref{crossover} (for
instance, we take $q=2$ and use the data set with $I=50$ mA). To
determine the value of the crossover, we introduce a $\Delta(s)$
function as:
\begin{eqnarray}
\Delta(s)=\sqrt{\left[F(s)-F_{\rm{Linear}}(s)\right]^2}
\label{chi_cross}
\end{eqnarray}
for each $q$, where $F(s)$ and $F_{\rm{Linear}}(s)$ are the
fluctuation functions for the original data and the filtered data
produced by the F-DFA method (see below), respectively. In Figure
\ref{chi_cross}, we plot $\Delta(s)$ as a function of $s$ for the
plasma fluctuations. Hereafter, for convenience we omit the
subscript $q$  and take $q=2$ unless expressed otherwise. The
crossover occurs at $s_{\times}\sim 0.02$ sec, corresponding to
$\nu_{\times}\sim 50$ sec$^{-1}$. Clearly, the fluctuation function
for $s\leq s_{\times}$ has the same scaling behavior as the noise
without trends (see Figure \ref{crossover}).

\begin{figure}[t]
\epsfxsize=9truecm\epsfbox{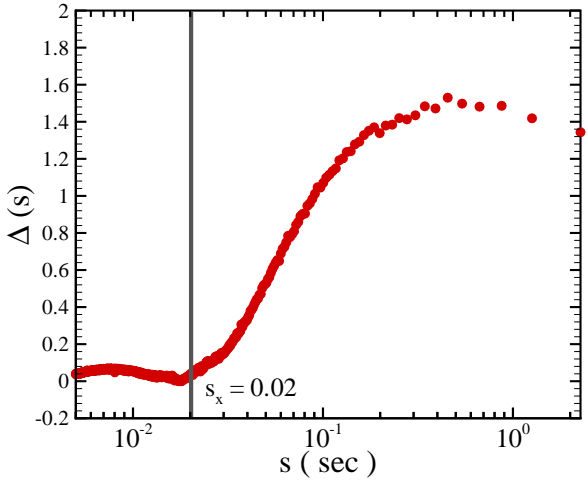} \epsfxsize=8truecm \narrowtext
\caption{The function $\Delta(s)$ for the plasma fluctuations versus
$s$.} \label{chi_cross}
\end{figure}
%\begin{figure}
%\begin{picture}(100,100)(0,0)
%\special{psfile=power_heavy_light.eps vscale=50 voffset=-250
%hoffset=200 hscale=50} \end{picture}
%\end{figure}

\begin{figure}[t]
\epsfxsize=9truecm\epsfbox{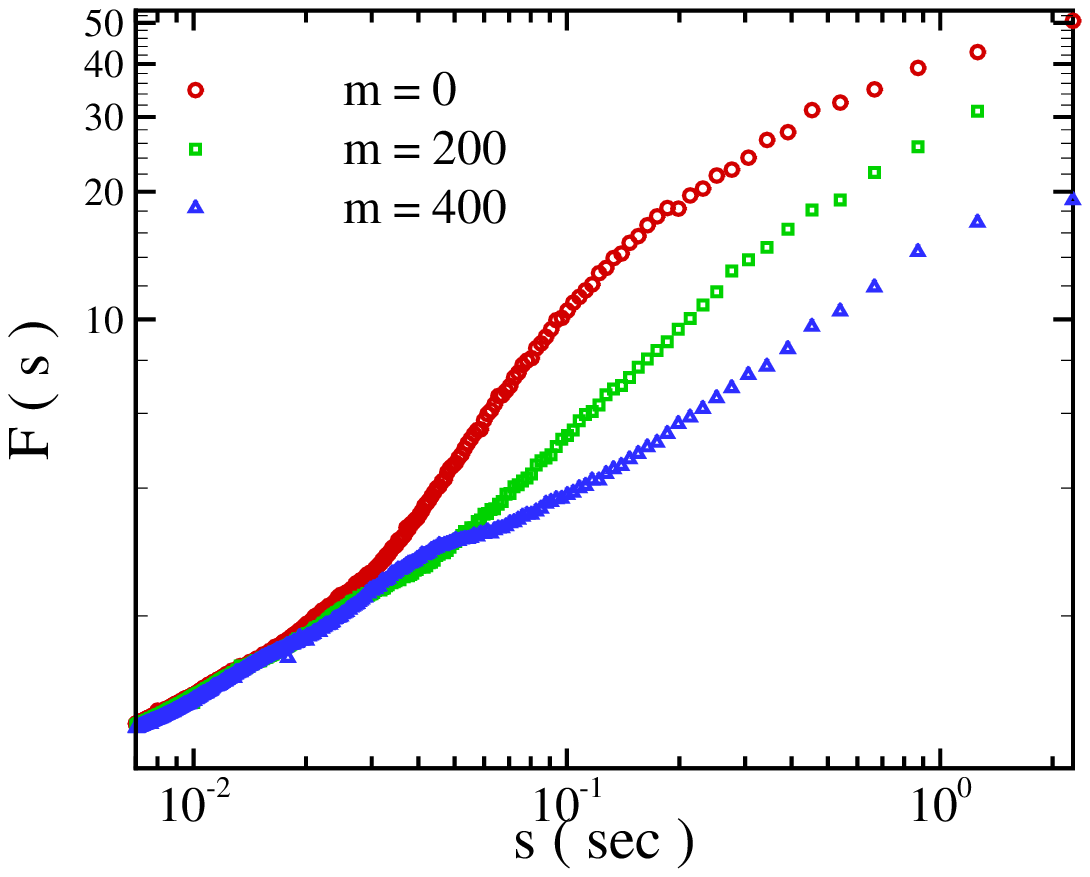} \narrowtext \caption{The MF-DFA
functions $F_q(s)$ in terms of the time scale $s$, in a log-log
plot. The original time series, $m=0$, the truncation of the first
200, $m=200$, and 400 terms, $m=400$. This plot is for a typical
value of the electrical discharge current intensity, $I=50$ mA.}
\label{truncate}
\end{figure}

\begin{figure}
\epsfxsize=9truecm\epsfbox{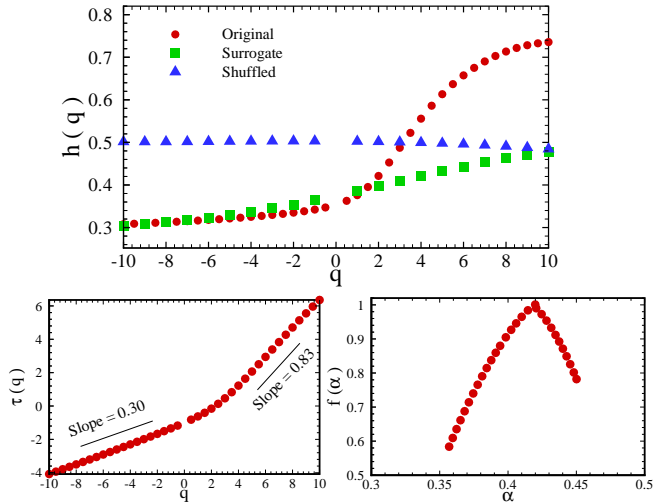} \narrowtext \caption{Upper
panel shows the generalized Hurst exponent versus $q$ for the
original, the surrogate, and the shuffled series without the
sinusoidal trend. Lower panel left and right correspond to the
classical multifractal scaling exponent and the singularity spectrum
$f(\alpha)$ for the data set at $I=50$ mA, respectively.}
\label{result1}
\end{figure}

As expressed in the last section, to cancel the sinusoidal trend in
the MF-DFA, we have applied the F-DFA method. Indeed, we truncate
the lowest frequencies up to the value that the regression of the
linear fitting of the corresponding log-log plot of the fluctuation
function for truncated data reaches 0.95. It is well known that if
we truncate the frequencies more than the necessary values for
eliminating the crossover, some statistical properties of the
underlying noise might be lost \cite{kunhu,trend2,chi05}. To
eliminate the crossover scales, we need to remove approximately the
first 400 terms of the Fourier expansion. Then, by the inverse
Fourier transformation, the fluctuations without the sinusoidal
trend are retrieved. The result is shown in Figure \ref{truncate}.
%According to Figure \ref{truncate}, for eliminating the crossover
%scales, we need to remove approximately the first 400 terms of the
%Fourier expansion. Then, by the inverse Fourier transformation, the
%fluctuations without the sinusoidal trend are retrieved.
The generalized Hurst exponent, the classical multifractal scaling
exponents, and the singularity spectrum for the data after the
elimination of sinusoidal trends are illustrated in Figure
\ref{result1}.
%for the data without sinusoidal trends are illustrated in Figures
%\ref{result1}.
In addition, we estimated the errors at $1\sigma$ confidence level
for all the derived values reported in all the tables, using the
common statistical method \cite{squ85}. The cleaned time series is a
multifractal process, as indicated by the strong $q$-dependence of
the generalized Hurst exponents \cite{bun02}. The $q$-dependence of
the classical multifractal scaling exponent $\tau(q)$ has different
behaviors for $q<0$ and $q>0$. For both positive and negative values
of $q$, the value of slopes of $\tau(q)$ are indicated in Figure
\ref{result1}. As mentioned before for $q<0$, small fluctuations
will be dominant in the fluctuation function, whereas for $q>0$ the
large fluctuations represent the dominant effect in $F_q(s)$. From
the statistical point of view, usually, for a multifractal
anti-correlated series, namely, one with $H<0.5$, the value of
$h(q)$ for $q<0$ is smaller than the generalized Hurst exponent for
positive moments. This is due to the fact that, the number of the
large fluctuations are statistically much more than the small
fluctuations in the time series. In other words, the series is
intermittent. The same circumstances arise for the plasma
fluctuations. In the presence of free charge, every large deviations
in the electrostatic equilibrium would be shielded by a cloud of
oppositely charged particles \cite{allan,chen,cap}. Therefore, we
expect the plasma fluctuations to be an anti-correlated series. The
value of the Hurst exponent confirms that the data set is a
stationary process. According to the MF-DFA results, all of the
discharge current series behave as weak anti-correlated process. The
value of the Hurst exponent and the classical multifractal exponent
for $q=2$, in the region that the sinusoidal trend is not
pronounced, calculated via the MF-DFA method are reported in Table
\ref{Tab1}. The correlation and power spectrum exponents are given
in Table \ref{Tab11}.

In spite of the power of the MF-DFA method, some cases encounter
problems and the method yields inaccurate results. The DFA method
can only determine a positive Hurst exponent, but yields an
inaccurate result for the strongly anti-correlated recorded data,
when $H\simeq 0$. To avoid this situation, one should use the
integrated data. In that case the series is the so-called double
profiled data set. The corresponding Hurst exponent is
$H=\bar{H}-1$, where $\bar{H}$ is derived from the DFA method for
the double profiled series \cite{sadeghsun,bun02}.

\begin{figure}[t]
\epsfxsize=9truecm\epsfbox{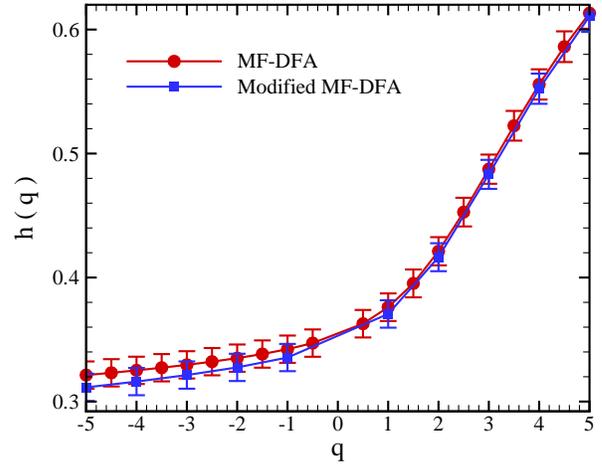} \narrowtext
\caption{Comparison between the modified and commonly used MF-DFA
for a typical value of the electrical-discharge current intensity,
$I=50$ mA.  } \label{modified}
 \end{figure}

Let us now discuss the finite size effect of the data set on the
results. According to the recent analysis in Ref.
\cite{physa,trend3,trend4}, a deviation from the DFA results occurs
in short records. The modified version of the MF-DFA should be used
for such cases \cite{physa,trend3,trend4}. Usually, in the MF-DFA
method, the deviation from a straight line in the log-log plot of
Eq. (\ref{Hq}) occurs for small scales $s$. The deviation limits the
capability of DFA for determining the correct correlation behavior
at very short scales, and in the regime of small $s$. The modified
MF-DFA is defined as follows \cite{physa,trend3,trend4}
\begin{eqnarray} F^{\rm mod}_q(s) & = & F_q(s) {\langle [F_q^{\rm shuf}(s')]^2 \rangle^{1/2}\, s^{1/2}
\over \langle [F_q^{\rm shuf}(s)]^2 \rangle^{1/2} \,
s'^{1/2} } \quad {\rm (for} \, s' \gg 1)\nonumber\\
\label{fmod}
\end{eqnarray}
where $\langle[F_q^{\rm shuf}(s)]^2\rangle^{1/2}$ denotes the usual
MF-DFA fluctuation function [defined in Eq. (\ref{fdef})], averaged
over several configurations of the shuffled data taken from the
original time series, and $s'\approx N/40$. The values of the Hurst
exponent obtained by the modified MF-DFA methods for the time series
are reported in Table \ref{mod}. The maximum relative deviation of
the Hurst exponent, which is computed by the modified MF-DFA,
relative to the MF-DFA for the original data, is approximately
2.7\%. Moreover, Figure \ref{modified} shows a comparison between
the generalized Hurst exponent, derived by the common MD-DFA1 and
modified MF-DFA. It indicates that the modified MF-DFA is consistent
with the MF-DFA at $1\sigma$ confidence level for various moments.

\begin{figure}[t]
\epsfxsize=9truecm\epsfbox{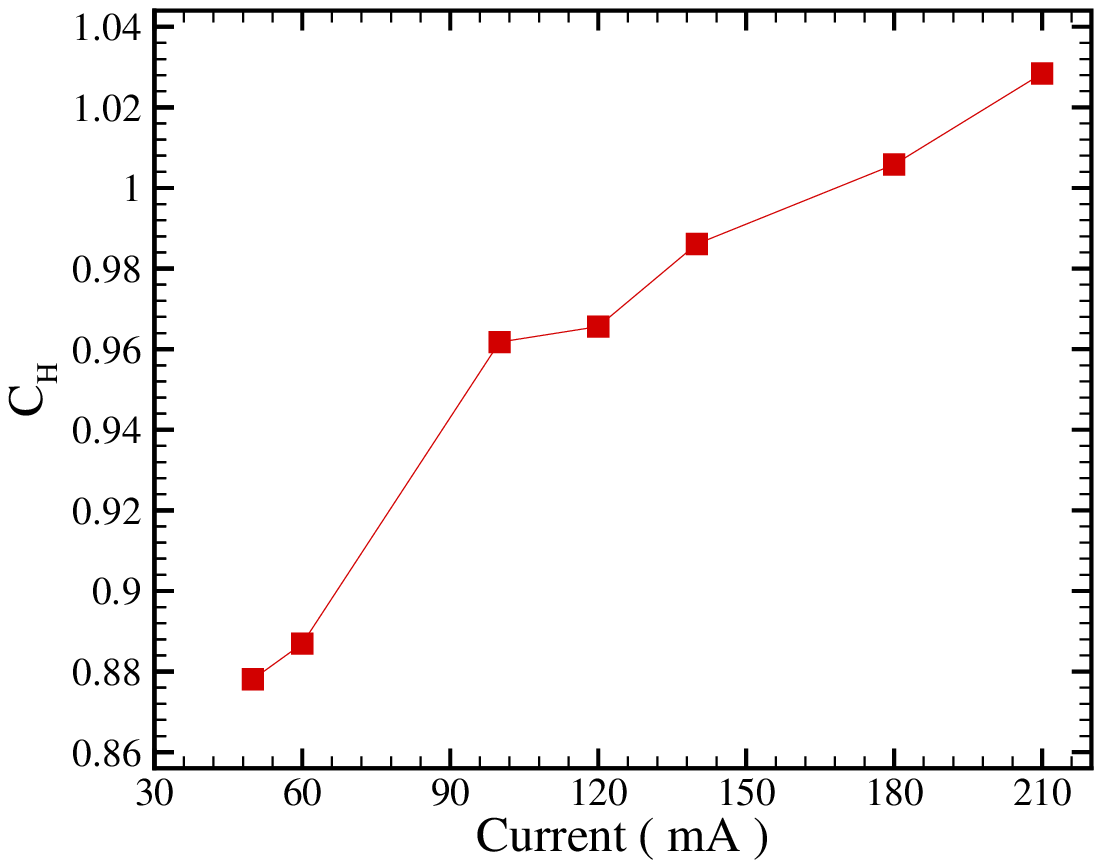} \narrowtext \caption{The value
of ${\mathcal{C}_{H}}$ as a function of the discharge current
intensity.} \label{intercept}
\end{figure}

By inspecting the log-log plot of the fluctuation function versus
$s$ (i.e., $F_2(s)={\mathcal{C}_{H}}s^{H}$), we find the dependence
of its amplitude on the electrical discharge current. Indeed, Ref.
\cite{sadeghriver} showed that this amplitude is given by
\begin{eqnarray}
{{\mathcal{C}}^2_{{\mathcal{H}}}}&=&\frac{\sigma^2}{(2H+1)}-\frac{4\sigma^2}{2H+2}
\nonumber\\
& & +3\sigma^2\left(\frac{2}{H+1}-\frac{1}{2H+1}\right)\nonumber\\
&&\qquad-\frac{3\sigma^2}{(H+1)}\left(1-\frac{1}{(H+1)(2H+1)}\right)
\label{a77}
\end{eqnarray}
where $\sigma^2=\left\langle x(i)^2\right\rangle$ is the variance of
the data. Figure \ref{intercept} shows the value of the amplitude as
a function of the current.

 \section{Search for origin of multifractality in the data set}

Due to the strong $q$ dependence of the generalized Hurst exponent
and the distinct slopes $\tau(q)$ for the various moments, the
remaining data set, after the elimination of the sinusoidal trends,
has multifractal properties. In this section we search for the
source of the multifractality. In general, two different types of
multifractality in time series may exist:

(1) Multifractality due to the fatness of the probability density
function (PDF) of the time series, in comparison to a Gaussian PDF.
In this case, the multifractality cannot be removed by shuffling the
series, because the correlations in the data set are affected by the
shuffling, while the PDF of the series is invariant.

(2) Multifractality due to different types of correlations in the
small and large scale fluctuations. In this case, the data may have
a PDF with finite moments, e.g., a Gaussian distribution. Thus, the
corresponding shuffled time series will exhibit monofractal scaling,
since all the correlations are destroyed by the shuffling procedure.
If both kinds of multifractality are present, the shuffled series
will exhibit weaker multifractality than the original series.

The easiest way to distinguish the type of multifractality is by
analyzing the corresponding shuffled and surrogate time series. The
shuffling of the time series destroys the correlation. Therefore, if
the multifractality only belongs to the correlation, we should find
$h_{\rm shuf}(q)=0.5$. The multifractality due to the fatness of the
PDF series is not affected by the shuffling procedure. On the other
hand, to determine whether the multifractality is due to the
broadness or fatness of the PDF, the surrogate data are used
\cite{theiler}. If multifractality in the time series is due to a
broad PDF, $h(q)$ for surrogate data will be independent of $q$. If
both kinds of multifractality are present in fluctuations, the
shuffled and surrogate series will exhibit weaker multifractality
than the original one. The utility of the two tests was first
demonstrated by Ivanov {\it et al.} \cite{6735,641,nature}.

To check the nature of the multifractality, the we compare the
fluctuation function $F_q(s)$ for the original series (after removal
of the sinusoidal trends) with the results for the corresponding
shuffled, $F_q^{\rm shuf}(s)$ and that of the surrogate series,
$F_q^{\rm sur}(s)$. The differences between the two fluctuation
functions with the original one, directly indicate the presence of
the correlations or broadness of the PDF in the original series. The
differences can be observed in plots of $F_q(s)/F_q^{\rm shuf}(s)$
and $F_q(s)/F_q^{\rm sur}(s)$ versus $s$ \cite{bun02}. Since the
anomalous scaling due to a broad PDF affects $F_q(s)$ and $F_q^{\rm
shuf}(s)$ in the same way, only multifractality due to the
correlations will be observed in $F_q(s)/F_q^{\rm shuf}(s)$. The
scaling behavior of the two ratios are given by
\begin{equation}
F_q(s)/F_q^{\rm shuf}(s) \sim s^{h(q)-h_{\rm shuf}(q)}=s^{h_{\rm
cor}(q)} \label{HqCor}
\end{equation}
\begin{equation}
F_q(s)/F_q^{\rm sur}(s) \sim s^{h(q)-h_{\rm sur}(q)}=s^{h_{\rm
PDF}(q)} \label{Hqpdf}
\end{equation}

If only the fatness of the PDF is responsible for the
multifractality, one should obtain, $h(q)=h_{\rm shuf}(q)$, and,
$h_{\rm cor}(q)=0$. On the other hand, deviations from $h_{\rm
cor}(q)=0$ indicates the presence of correlations, while the $q$
dependence of $h_{\rm cor}(q)$ indicates that multifractality is due
to the correlation. If both the distribution and correlation
multifractality are present, both $h_{\rm shuf}(q)$ and $h_{\rm
sur}(q)$ will depend on $q$.

\begin{figure}[t]
\epsfxsize=9.truecm\epsfbox{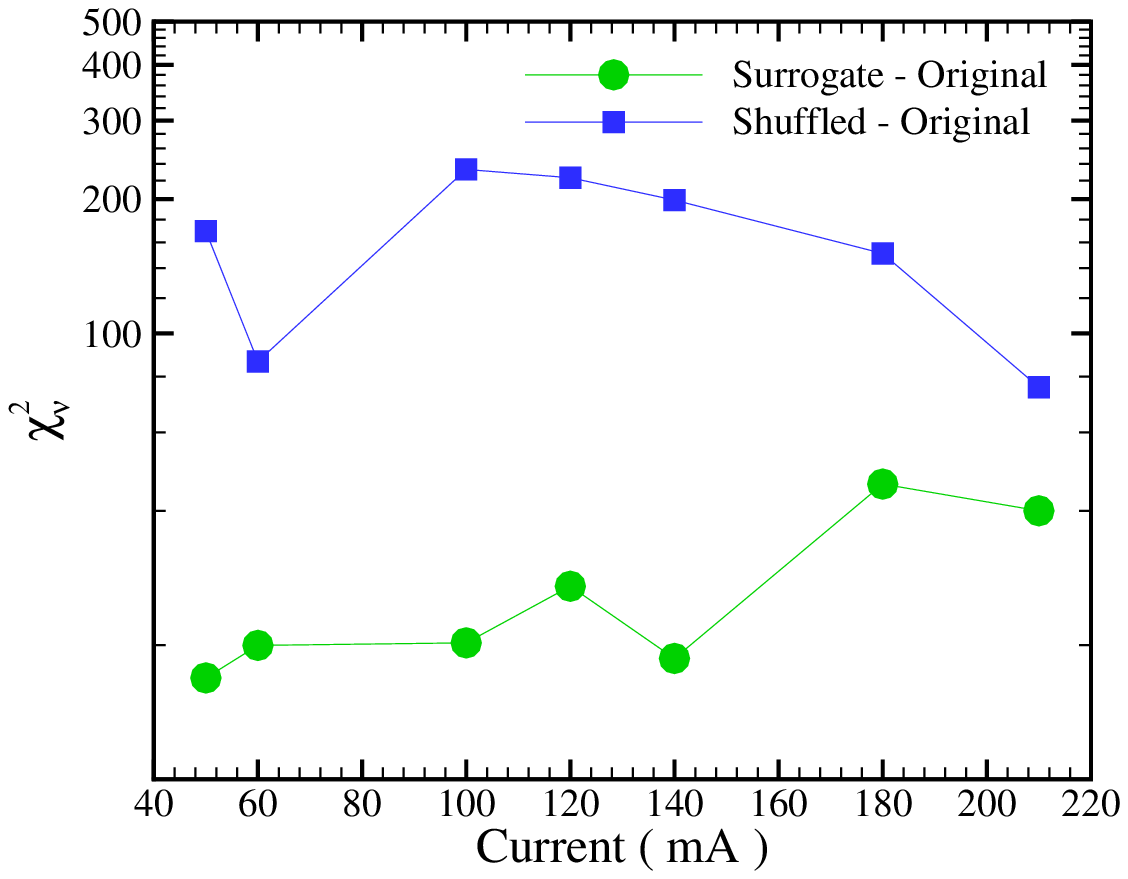} \narrowtext
\epsfxsize=9.truecm\epsfbox{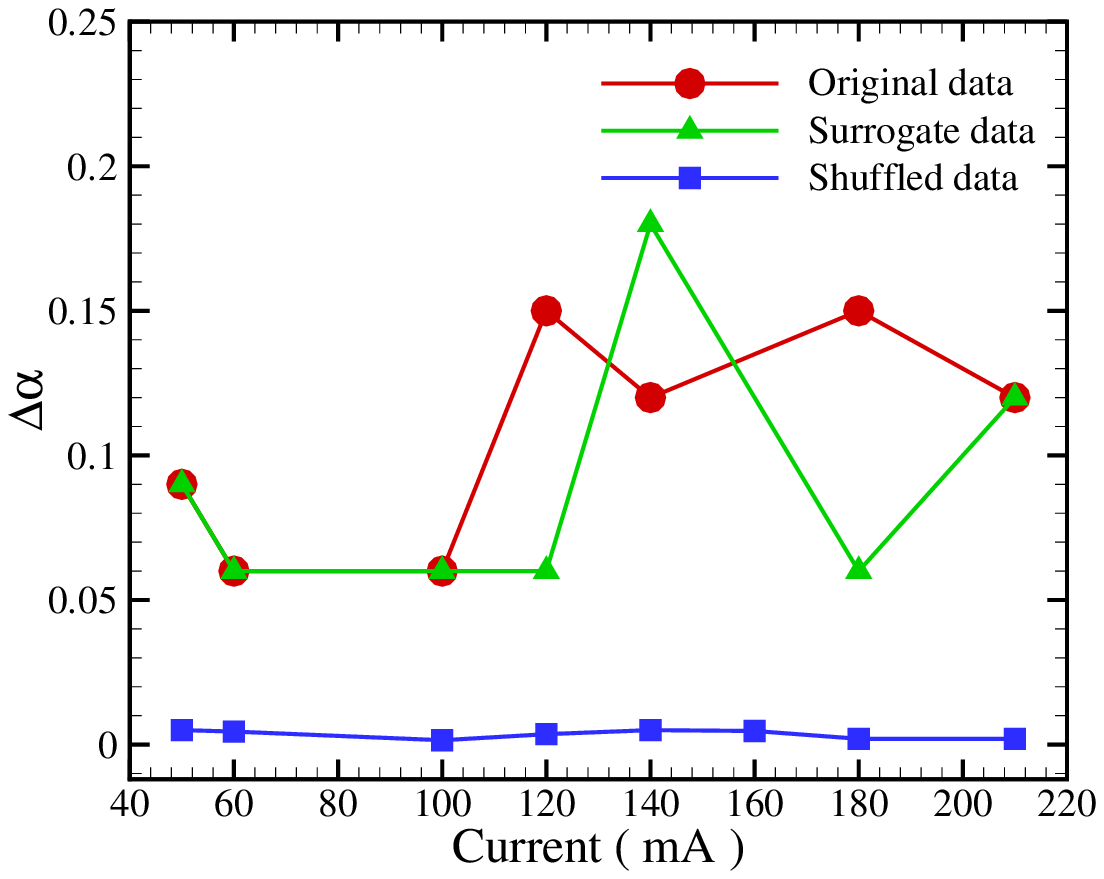} \narrowtext \caption{Upper
panel shows to the values of $\chi^2_{\nu}$ for the data sets. Lower
panel shows the width of the singularity strength, $\Delta\alpha$,
for the original, surrogate and shuffled data sets in various
electrical discharge currents, obtained by the MF-DFA.}
\label{alphachi}
\end{figure}

The $q$ dependence of the exponent $h(q)$ for the original,
surrogate, and shuffled time series are shown in Figures
\ref{result1}. The $q$ dependence of  $h_{\rm PDF}$ shows that the
multifractality nature of the time series is almost due to the
correlation. However, the value of $h_{\rm PDF}(q)$ is deviated from
zero which confirms the multifractality due to the broadness of the
PDF is much weaker than the multifractality due to the correlation.
The deviation of $h_{\rm sur}(q)$ and $h_{\rm shuf}(q)$ from $h(q)$
may be determined by using the $\chi^2$ test:

\begin{equation}
\chi^2_{\diamond}=\sum_{i=1}^{N}\frac{[h(q_i)-h_{\diamond}(q_i)]^2}{\sigma(q_i)^2+\sigma_{\diamond}(q_i)^2}
\label{khi}
\end{equation}
where the symbol $\diamond$ can be replaced by "sur" and "shuf", in
order to determine the confidence level of $h_{\rm sur}$ and $h_{\rm
shuf}$, the generalized Hurst exponents of the original series,
respectively. The value of the reduced chi-square,
$\chi^2_{\nu}=\chi^2/{\mathcal{N}}$ where ${\mathcal{N}}$ is the
number of degree of freedom, for the shuffled and surrogate time
series are shown in the upper panel of Figure \ref{alphachi}. On the
other hand, the lower panel of Figure \ref{alphachi} illustrates the
width of the singularity spectrum, $f(\alpha)$, i.e., $\Delta
\alpha=\alpha(q_{min})-\alpha(q_{max})$, for the original,
surrogate, and shuffled data set. These values also confirm that the
multifractality due to the correlations is dominant \cite{paw05}.

The values of the generalized Hurst exponent $h(q=2.0)$, and the
multifractal scaling $\tau(q=2)$ for the original, shuffled, and
surrogate of the discharge fluctuation obtained with the MF-DFA
method are reported in Table \ref{Tab1}.

\begin{table}[htp]
\caption{\label{Tab1}Values of $H=h(q=2)$ and the classical
multifractal scaling exponents for $q=2.0$ for the original,
surrogate and shuffled data set and different electrical currents,
obtained by the MF-DFA.}
\begin{center}
\begin{tabular}{|c|c|c|c|}
  Sample& & $H$ & $\tau$  \\ \hline
     &{\rm Original} & $0.42\pm 0.01$ &$-0.16\pm0.02$      \\\hline
   $50{\rm mA}$ &{\rm Surrogate}&$0.40\pm0.01$ &$-0.21\pm0.02$     \\\hline
  &{\rm Shuffled} & $0.50\pm0.01$ &$0.004\pm0.020$  \\ \hline\hline

       &{\rm Original} & $0.45\pm 0.01$ &$-0.10\pm0.02$    \\\hline
   $60{\rm mA}$ &{\rm Surrogate}&$0.42\pm0.01$ &$-0.16\pm0.02$     \\\hline
  &{\rm Shuffled} & $0.50\pm0.01$ &$0.002\pm0.020$  \\ \hline\hline

     &{\rm Original} & $0.37\pm 0.01$ &$-0.25\pm0.02$    \\\hline
   $100{\rm mA}$ &{\rm Surrogate}&$0.36\pm0.01$ &$-0.28\pm0.02$      \\\hline
  &{\rm Shuffled} & $0.49\pm0.01$ &$0.00\pm0.02$    \\ \hline\hline

     &{\rm Original} & $0.38\pm 0.01$ &$-0.23\pm0.02$     \\\hline
   $120{\rm mA}$ &{\rm Surrogate}&$0.36\pm0.01$ &$-0.28\pm0.02$    \\\hline
  &{\rm Shuffled} & $0.50\pm0.01$ &$0.00\pm0.02$  \\ \hline\hline

     &{\rm Original} & $0.41\pm 0.01$ &$-0.17\pm0.02$     \\\hline
   $140{\rm mA}$ &{\rm Surrogate}&$0.40\pm0.01$ &$-0.21\pm0.02$      \\\hline
  &{\rm Shuffled} & $0.50\pm0.01$ &$0.00\pm0.02$  \\ \hline\hline

  %   &{\rm Original} & $0.59\pm 0.01$ &$0.17\pm0.02$  \\\hline
 %  $160{\rm mA}$ &{\rm Surrogate}&$0.53\pm0.01$ &$0.07\pm0.02$ \\\hline
 % &{\rm Shuffled} & $0.49\pm0.01$ &$-0.01\pm0.02$ \\ \hline\hline

     &{\rm Original} & $0.45\pm 0.01$ &$-0.09\pm0.02$   \\\hline
   $180{\rm mA}$ &{\rm Surrogate}&$0.40\pm0.01$ &$-0.20\pm0.02$     \\\hline
  &{\rm Shuffled} & $0.50\pm0.01$ &$-0.003\pm0.020$   \\ \hline\hline

     &{\rm Original} & $0.48\pm 0.01$ &$-0.04\pm0.02$    \\\hline
   $210{\rm mA}$ &{\rm Surrogate}&$0.44\pm0.01$ &$-0.13\pm0.02$   \\\hline
  &{\rm Shuffled} & $0.51\pm0.01$ &$0.01\pm0.02$
\end{tabular}
\end{center}
\end{table}

\begin{table}[htp]
\caption{\label{Tab11} Values of the correlation and power spectrum
exponents for the original data set with different electrical
currents, obtained by the MF-DFA.}
\begin{center}
\begin{tabular}{|c|c|c|}
Sample& $\gamma$ & $\beta$  \\ \hline $50{\rm mA}$  & $1.16\pm 0.02$
&$-0.16\pm0.02$       \\ \hline $60{\rm mA}$  & $1.10\pm 0.02$
&$-0.10\pm0.02$       \\ \hline $100{\rm mA}$  & $1.26\pm 0.02$
&$-0.26\pm0.02$      \\ \hline $120{\rm mA}$  & $1.24\pm 0.02$
&$-0.24\pm0.02$      \\ \hline $140{\rm mA}$  & $1.18\pm 0.02$
&$-0.18\pm0.02$      \\ \hline
% $160{\rm mA}$  & $0.82\pm 0.02$ &$0.18\pm0.02$     \\ \hline
$180{\rm mA}$  & $0.10\pm 0.02$ &$-0.10\pm0.02$      \\ \hline
$210{\rm mA}$  & $1.04\pm 0.02$ &$-0.04\pm0.02$
\end{tabular}
\end{center}
\end{table}

\begin{table}[htp]
\caption{\label{mod} Values of Hurst exponent using the regular
MF-DFA and modified MF-DFA for data sets.}
\begin{center}
\begin{tabular}{|c|c|c|}
Sample &  $ H$ & $H_{\rm Mod}$ \\ \hline $50{\rm mA}$ &$0.42\pm0.01$
&$0.42\pm0.01$     \\ \hline $60{\rm mA}$ &$0.45\pm0.01$
&$0.45\pm0.01$     \\ \hline $100{\rm mA}$ &$0.37\pm0.01$
&$0.38\pm0.01$     \\ \hline $120{\rm mA}$ &$0.38\pm0.01$
&$0.37\pm0.01$     \\ \hline $140{\rm mA}$ &$0.41\pm0.01$
&$0.42\pm0.01$     \\ \hline
% $160{\rm mA}$ &$0.59\pm0.01$ &$0.58\pm0.01$\\ \hline
180{\rm mA} & $0.45\pm0.01$ & $0.46\pm0.01$     \\ \hline
210{\rm  mA} & $0.48\pm0.01$ & $0.48\pm0.01$    %\\\hline
\end{tabular}
\end{center}
\end{table}

\section{Conclusion}

Discharge current fluctuations in plasma are affected by many
factors. From a statistical point of view, and in order to
understand the complexity of the fluctuations, we applied a robust
method, such as the detrending fractal analysis to infer the
complexity and multifractal features of the underlying plasma
fluctuations. In the presence of nonstationarity, non-detrending
methods will be encountered with some challenges, such that they
yield wrong or at least inaccurate results. Multifractal detrended
fluctuation analysis is well established for investigating noisy
time series, and can be used to gain deeper insight into the
processes that occur in nonstationary dynamical systems, such as
electrical discharge current.

We showed that the MF-DFA results for the time series for various
electrical currents have one crossover time scale, $s_{\times}$.
This crossover time scale is about $s_\times\sim 0.02$ second, and
is related to the sinusoidal trends. The crossover time scale which
discriminates the noise and trends intervals can be potentially
related to the coherent time scale in turbulent plasma. Plasma
fluctuations are not affected by external factors within time scale. %These
%crossovers indicate as a criteria for transition between one scale
%over which intrinsic fluctuation has dominant effect to other scale
%over which trends is dominant. In other words, crossover time scale
%which discriminates the noise and trends intervals could be
%potentially related to the coherent time scale in plasma turbulent
%in which plasma fluctuations are not affected by external factors.
To minimize the effect of trends and produce clean data set for
further investigation, we applied the Fourier detrended fluctuation
analysis to the data sets. Indeed, after applying the F-DFA, the
data set without sinusoidal trends is recovered, and the spurious
behavior in the MD-DFA results disappear. Applying the MF-DFA method
on the cleaned data set demonstrated that the discharge current
fluctuations are stationary time series.

According to the value of the Hurst exponent, computed by the MF-DFA
method, all the discharge current time series behave as weak
anti-persistent processes. These findings can be interpreted as
follows: in the presence of free charges, every large deviation from
the electrostatic equilibrium is shielded by a cloud of
oppositely-charged particles \cite{allan,chen,cap}. This also may be
related to the fast dissipation of turbulent kinetic energy in
plasma. Our results also confirmed that the multifractality nature
is a global property of various plasma data based on different
experimental setups \cite{bud1}. We found that Hurst exponent and
multifractality nature based on singularity spectrum didn't depend
on the discharge current intensity (see Table \ref{mod} and Figure
\ref{alphachi}). This result indicates that increasing the amount of
charged particles at least in our experimental setup almost don't
alter the statistical properties of the plasma fluid. But it is
interesting to extend these analysis to a broader set of plasma data
in various working pressure and check their statistical properties.
The $q$-dependence of $h(q)$ and $\tau(q)$ indicated that the data
sets have multifractal properties.

%The absence of any dependency of Hurst exponent and multifractality
%nature based on singularity spectrum to the discharge current (see
%Table \ref{mod} and Figure \ref{alphachi}) indicates that increasing
%the amount of charged particles at least in our experimental setup
%almost don't affect on statistical properties of plasma fluid. But
%it is interesting to extend this analysis to a broader set of plasma
%data in various working pressure and check the dependency of their
%statistical properties.

The value of $h(q)$ for $q>0$ is larger than the same quantity for
$q<0$, indicates that the number of large fluctuations are
statistically larger than the small fluctuations in the time series.
Our results show that, amplitude of the fluctuation function
${\mathcal{C}_{H}}$ is a monotonous function in terms of discharge
current intensity. This demonstrate that by increasing the current
intensity, plasma instability will be occurred \cite{wei} and
consequently one has large variance for data set (see Figure
\ref{intercept}).

In order to recognize the nature of the multifractality, we compared
the generalized Hurst exponent of the original time series with
those of the shuffled and surrogate ones. The comparison indicated
that the multifractality due to the correlations makes more
significant contribution than the broadness of the probability
density function of the current fluctuations.

{\bf Acknowledgements} The authors are grateful to the anonymous
referee for his/her useful comments. The experimental part of this
research was supported by Tabriz University. We also thank M. Sahimi
for a critical reading of the manuscript.

\end{document}